# COMPARISON BETWEEN CONVENTIONAL LOAD FLOW, QSTS SIMULATION, AND DYNAMIC SIMULATION TO ASSESS THE OPERATION OF STEP VOLTAGE REGULATORS IN ACTIVE DISTRIBUTION NETWORKS

**V.M. SOUZA***
Federal University of Pará
Brazil

**H.R. BRITO**
Federal University of Pará
Brazil

**J.P.A. VIEIRA**
Federal University of Pará
Brazil

**M.E.L. TOSTES**
Federal University of Pará
Brazil

**U.H. BEZERRA**
Federal University of Pará
Brazil

**H.N.S. CARDOSO**
Equatorial Energia Utility Company
Brazil

**M.S. COSTA**
Equatorial Energia Utility Company
Brazil

*Abstract* – The assessment of step voltage regulator (SVR) operation in active distribution networks requires computational analysis tools capable of tackling the emerging technical challenges. Conventional load flow (CLF), quasi-static time series (QSTS) and dynamic simulations are typically employed to investigate high-penetration distributed generation (DG) interconnection impacts. Regarding the SVR runaway condition phenomenon, however, a consensus has yet to be reached on the most cost-effective simulation technique for capturing and reproducing the correct sequence of events. This work presents a comparative study of the CLF, QSTS and dynamic simulation techniques through modelling and analysis of two SVR-controlled test-feeders, in order to evaluate each approach's performance in addressing scenarios of DG-caused reverse active power flow. Detailed descriptions of feeder voltage profile and SVR tap operations are provided to facilitate understanding of the mechanisms that characterize SVR runaway condition, as well as the advantages and drawbacks of each of the studied simulation techniques.

*Keywords:* Step Voltage Regulator – Distribution Network – Conventional Load Flow – QSTS Simulation – Dynamic Simulation – Distributed Generation – Runaway Condition

## 1 INTRODUCTION

The increasing presence of distributed generation (DG) units in low and medium-voltage distribution grids has motivated the development of analysis and planning tools capable of tackling the arising operational challenges. One of the main impacts stemming from the growing number of DG-grid interconnections requested by independent power producers (IPPs) is the feeder overvoltage problem, further aggravated by the high R/X cable ratio typically found at the distribution level [1]. In this context, the adequate operation of traditional voltage control devices such as the step voltage regulator (SVR), frequently employed in Brazil at long rural distribution feeders, is paramount for maintaining steady-state line voltages within acceptable operating ranges, as per local regulatory standards.

However, high DG penetration levels can adversely affect SVR performance, depending on both devices' control mode settings. The most common negative consequences of such association include excessive SVR tap operations, when the DG has a primary energy source of high output variability, and loss of SVR voltage control capability due to runaway condition, when the DG power supply exceeds the total feeder demand downstream of the SVR location [2]-[3]. Both cases entail excessive SVR wear and tear, reduction of the device's useful life and higher maintenance costs. In addition, overvoltage or undervoltage issues are possible outcomes, especially on the lower short-circuit level side of the feeder.

* valeria.monteiro12@gmail.com

Many authors opt to employ traditional simulation techniques when studying the aforementioned DG integration impacts, usually resorting to either conventional load flow (CLF) methods [4]-[5] or real-time dynamic simulations [6]-[7]. As an alternative, quasi-static time series (QSTS) analyses have found wide application in the assessment of SVR-controlled distribution systems subjected to high levels of photovoltaic generation, so as to better understand high output variability impacts in a time window of hours, days or even months. Such investigations attest to the effectiveness of the QSTS method and highlight its advantages with respect to traditional approaches [8]-[10].

On the other hand, proper reproduction of SVR runaway condition characteristics – from the moment the active power flow through the device is reversed until the resulting sequence of tap operations is concluded – requires a significantly smaller time window, in the order of seconds or minutes. For this reason, in regard to the runaway phenomenon, the applicability of the QSTS method and its performance vis-à-vis traditional simulation techniques are not evident. Refs. [3] and [11] present SVR runaway condition mitigation strategies based on QSTS simulations, but no other techniques are addressed. Ref. [12] compares the CLF and QSTS methods in terms of chronological accuracy, but does not include dynamic simulations, useful frames of reference for their level of detail, in its results. It is noticeable that a consensus has yet to be reached on the most cost-effective simulation technique for SVR runaway condition studies.

This work aims to conduct a comparative study of the CLF, QSTS and dynamic simulation techniques with special focus on their accuracy in reproducing the sequence of events that leads to SVR runaway condition in the presence of a high-penetration DG unit. CLF and QSTS analyses are performed in the Open Distribution Simulator Software (OpenDSS), whereas dynamic analyses are performed in the Analysis of Electro-mechanical Transients Software (ANATEM). Two distribution test-feeders are modelled in both simulation environments using equivalent system parameters and operating conditions consistent with the scope of the study. Comparative results of feeder voltage profile and SVR tap operations are used to assess the voltage regulating mechanisms, as well as the advantages and drawbacks of each of the simulation techniques.

## 2 OPERATIONAL ASPECTS OF STEP VOLTAGE REGULATORS

The SVR device typically employed in Brazilian medium-voltage distribution feeders is essentially an autotransformer with a load tap changing mechanism in its series winding. Depending on feeder loading, the voltage induced at such winding can either be added to or subtracted from the primary voltage, which allows for bilateral adjustments of small deviations with respect to a user-defined voltage setpoint. Many SVR models adopt a total regulation range of ±10 %, equally divided into 33 discrete steps: 16 voltage raising positions, 16 voltage lowering positions, and the neutral position. Therefore, each tap operation amounts to a 0.625 %, or 0.00625 normalized p.u., change per step. Details concerning constructive aspects of the single-phase SVR device used in simulations are available in [3].

Fig. 1 shows the simplified dynamic model of the SVR. The input voltage ($V_{in}$) measured at the SVR regulation point is compared to the voltage setpoint ($V_{ref}$), thus resulting in a voltage error signal ($V_{error}$) sent to the "Measuring Element" block. The error is in turn compared to the deadband (D), an adjustable range of allowed variance around $V_{ref}$, and at times to the hysteresis band ($\varepsilon$), a parameter that mitigates frequent tap operations during temporary oscillations around D. If $V_{error}$ exceeds such limits, an activation signal ($V_{act}$) is sent to the "Tap Changer" block, triggering its timer relay. Although time delay schemes vary widely depending on Brazilian utility practices, many employ a double-time delay scheme, where the first tap operation trigger ($T_1$) is slower than the subsequent ones ($T_2 = T_3 = \ldots$). Once the relay times out, the tap changing command is sent to a motor drive unit, which mechanically carries out the operation. This procedure is repeated as many times as necessary until $V_{error}$ is within deadband limits.

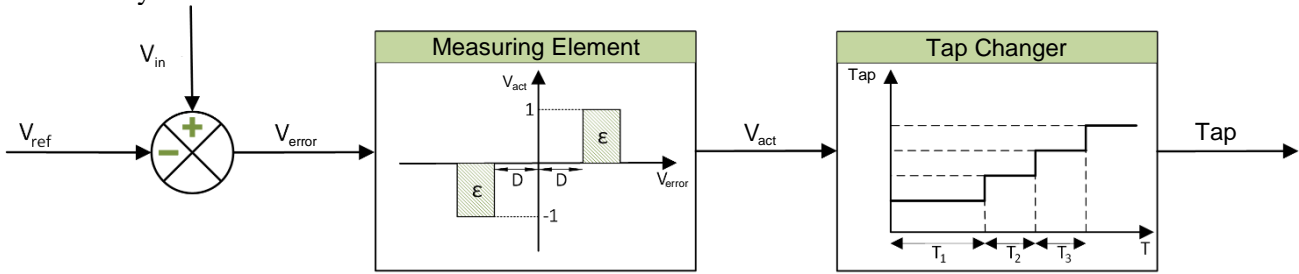

Fig. 1. Simplified dynamic model of a step voltage regulator.

The block diagram of Fig. 1 represents the SVR model developed in ANATEM for dynamic simulations. Its parameters, however, are also integrated to the iterative load flow calculations of CLF and QSTS simulations conducted in OpenDSS. In both platforms, SVR operation within the feeder is set to bidirectional mode, which is the recommended control setting when network reconfiguration via relay switching schemes is possible [11]. In addition, this setting is the most susceptible to the occurrence of DG-caused SVR runaway condition, especially when the DG is unable to alleviate system voltage deviations through reactive power exchanges, i.e., when it is set to operate in unity power factor control mode [12].

In bidirectional mode, the SVR determines its regulation point based on the direction of the active power flow. Fig. 2 depicts the situation of direct active power flow through the feeder, when the DG supplies less real power than the load center downstream of the SVR demands. In this case, the resulting active power flows through the SVR from the primary substation (PS) to the DG unit, and the device regulates point 2, located on the lower short-circuit level side. This operational scenario is considered acceptable since voltage control through tap changing is effective.

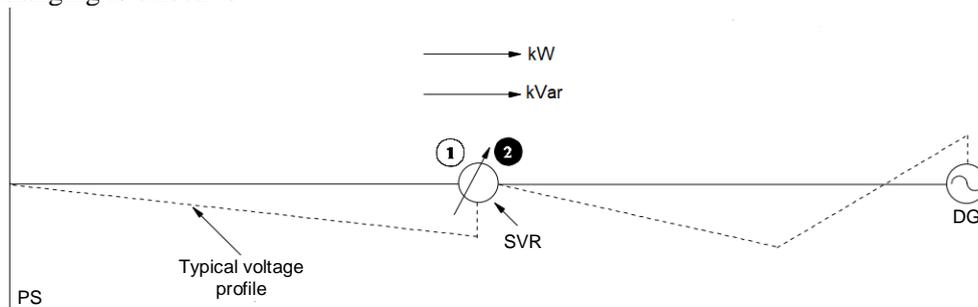

Fig. 2. Bidirectional SVR during direct active power flow. Source: adapted from [7].

Fig. 3 depicts the situation of reverse active power flow through the feeder, when the real power supplied by the DG exceeds the load center demand. In this case, the resulting active power through the SVR flows from DG to PS, and the device regulates point 1, located on the higher short-circuit level side. The tap changer thus operates in an effort to reduce the voltage at this point with negligible results, given the electrically strong nature of the PS side of the feeder. The outcome is a sequence of failed regulation attempts and, due to the ensuing reactive power flow, a net effect of significant voltage rise at point 2. Successive operations continue until the tap limit is reached, leading to a 10 % overvoltage downstream of the SVR.

The loss of SVR voltage control capability by attempting to regulate a point on the higher short-circuit level side of the feeder characterizes the phenomenon known in literature as reverse power tap changer runaway condition. The possibility of reverse active power flow through the SVR due to high DG penetration levels, even if temporarily, makes SVR bidirectional mode unacceptable in a real scenario. Besides the mentioned voltage violations, adverse effects include shortened SVR lifespan, excessive wear and tear and higher maintenance costs. These negative consequences justify the need for computational analysis tools that ensure proper understanding of the chronological progression of the runaway phenomenon.

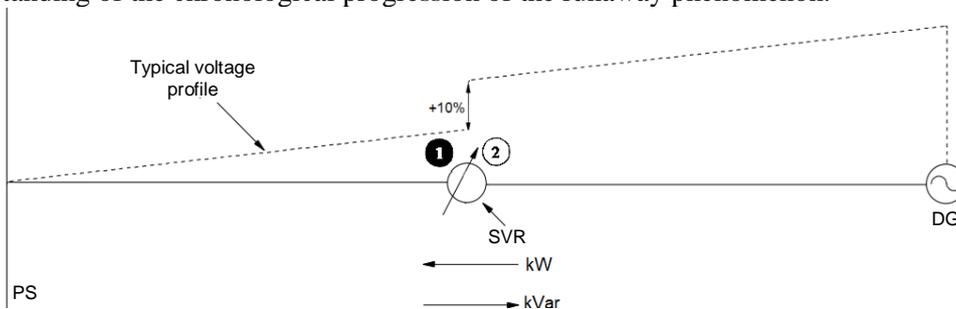

Fig. 3. Bidirectional SVR during reverse active power flow. Source: adapted from [7].

## 3 COMPUTATIONAL ANALYSIS TECHNIQUES

Fig. 4 compares required grid parameters for CLF, QSTS and dynamic simulations. It is evident that, among the three, the CLF method has the smallest number of requirements, as it only depends on feeder loading (active and reactive power demand) and DG specifications (active power injection and power factor (PF)). Thus, the SVR power flow is calculated iteratively and temporal dependencies do not influence convergence.

Iterative load flow solutions are also inherent to the QSTS formulation, although the method includes both time-varying parameters, such as load demand and DG penetration profiles, and time-dependent parameters, such as the power flow through the SVR based on its tap positions. With respect to DG-SVR interactions, the QSTS simulation is an asset for combining the simplicity of iterative calculations with the chronological accuracy of temporal considerations. Furthermore, this is accomplished without resorting to an unnecessary level of detail for low-frequency dynamics events, such as SVR tap operations [9].

The dynamic simulation, on the other hand, employs numerical integration techniques and requires not only load and generation profiles but also detailed modelling of network components such as the SVR and the DG. Despite the higher computational effort, it is the closest available approximation to a real impact study scenario, being an useful frame of reference for evaluating the performance of the iterative methods.

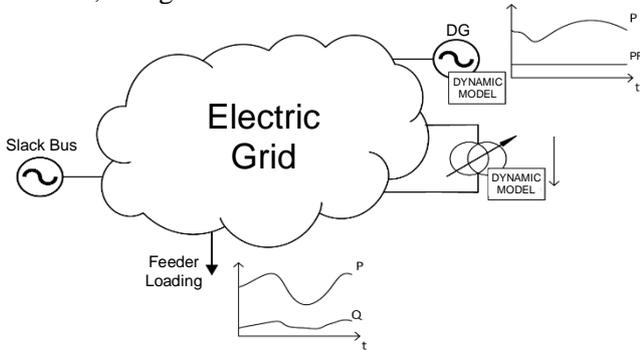
Fig. 4–(a). Grid parameters for dynamic simulation.

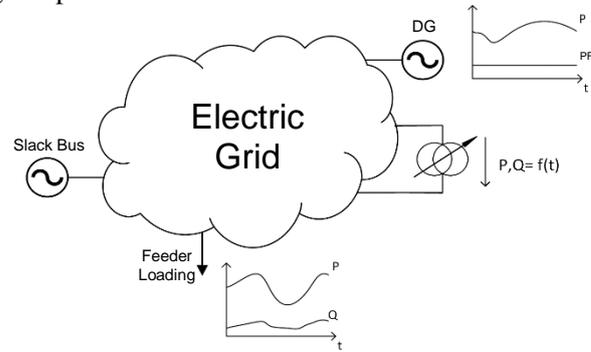
Fig. 4–(b). Grid parameters for QSTS simulation.

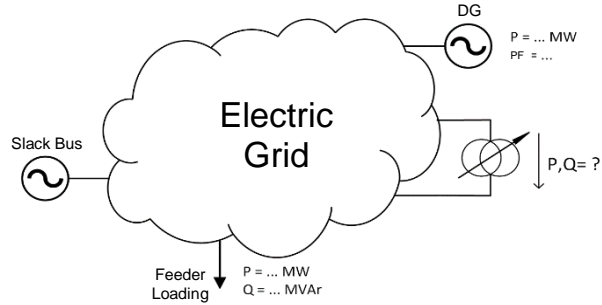
Fig. 4–(c). Grid parameters for CLF simulation.

## 4  RESULTS AND DISCUSSIONS

The comparison between the aforementioned analysis techniques is carried out using two test-feeders, both modelled in the OpenDSS and ANATEM programming environments. CLF and QSTS simulations are performed in OpenDSS, whereas dynamic simulations are performed in ANATEM.

In all scenarios, the SVR is set to operate in bidirectional mode with a deadband of 1 % (D = 1 %) around its voltage setpoint and with no hysteresis band ($\varepsilon = 0$). Additionally, a double-time delay scheme is adopted: first tap change delay of 30 s ($T_1 = 30$ s) and subsequent tap change delay of 5 s ($T_2 = T_3 = \ldots = 5$ s).

The time step for OpenDSS simulations is of 1 s, whereas the integration step size for ANATEM simulations is of 0.1 s. All cases studied have the same total simulation time of 350 s and are run on a machine with an Intel i7-6500U 2.5/2.6 GHz processor and 8 GB RAM. The reason for such choices is to facilitate as much as possible the comparative study between distinct platforms without compromising the validity of the results.

### 4.1  4-Bus Test-Feeder

Fig. 5 shows the first test-feeder: a simple 4-bus radial distribution network with 13.8 kV base voltage. The PS bus (B1) is the system's angular reference with 1.03 p.u. voltage setpoint. The SVR voltage setpoint is of 1 p.u. ($V_{ref} = 1$ p.u.). The DG unit is set to operate in unity power factor control mode and static ZIP parameters are assigned to the aggregated load model accordingly.

Fig. 6 shows the active power flow variation through the SVR as the DG real power injection increases in linear fashion. Its ramp-up behavior starts at 10 s of simulation and stabilizes after reaching 2.5 MW at 200 s. Once DG penetration exceeds feeder loading, reverse active power flow through the SVR occurs (at around 110 s). This changes henceforth the regulation point of the bidirectional SVR from bus B3 to bus B2.

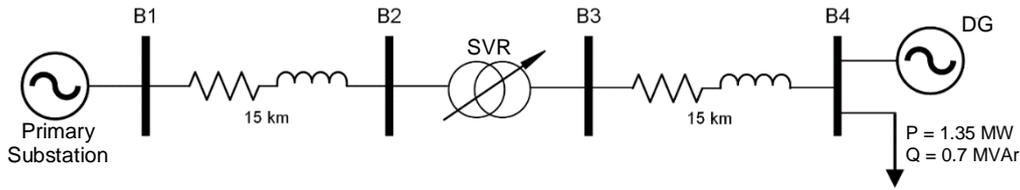

Fig. 5. Single-line diagram of the 4-bus test-feeder.

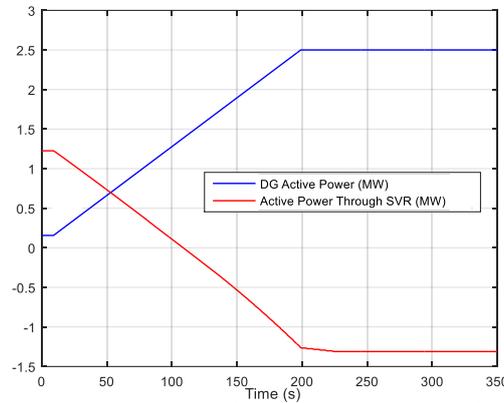

Fig. 6. DG active power ramp and active power flow through the SVR.

Fig. 7 compares the analysis techniques in terms of SVR tap evolution and voltage profiles of SVR-regulated buses. Results obtained from QSTS and dynamics simulations are remarkably similar. During direct active power flow, three tap operations for bus B3 voltage correction occur practically at the same moment in both simulations. During reverse active power flow, similarities are also noticeable since sequential tap changes closely coincide. The biggest difference lies in bus B3's magnitude of overvoltage – the outcome of failed attempts of regulating bus B2 voltage via SVR tap operations – which stabilizes at 1.1705 p.u. in the dynamic approach and at 1.1615 p.u. in the QSTS approach. The latter method, thus, presents a bus B3 steady-state voltage underestimation error of 0.77 %.

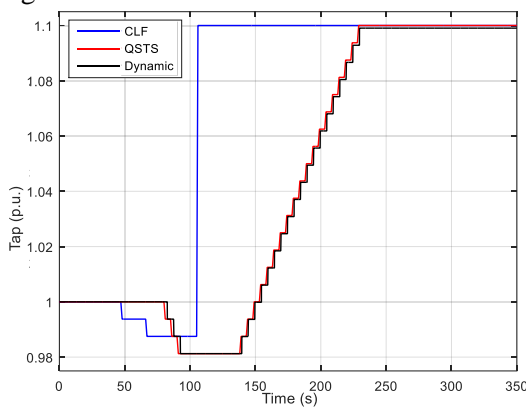
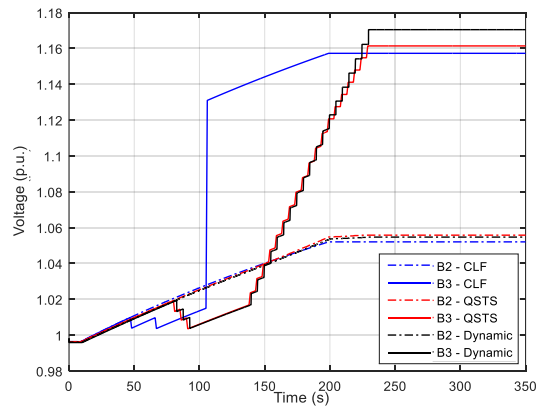

Fig. 7–(a). SVR tap evolution.  Fig. 7–(b). Voltage profiles of SVR-regulated buses.

Conversely, results obtained from CLF simulations differ in several aspects by not considering time-delay schemes associated with SVR tap operations, thereby leading to an inaccurate reproduction of the sequence of events. During direct active power flow, two tap changes, instead of three, occur without delay once bus B3 voltage violates the adopted deadband, and long before any control response in the other simulations. During reverse active power flow, SVR runaway condition is represented by an almost instantaneous tap excursion to the upper tap limit, resulting in overvoltage issues downstream of the device which are further aggravated by the DG ramp. In the CLF approach, bus B3 voltage stabilizes at 1.1573 p.u., indicating an underestimation error of 1.12% with respect to the dynamic approach frame of reference.

Fig. 8 shows complete feeder voltage profiles for each technique. Bus B2 voltage is unaffected by SVR tap changes, whereas buses B3 and B4 develop severe overvoltage. In terms of chronological accuracy, only the QSTS and dynamics approaches provide suitable results for a reliable assessment of SVR runaway condition.

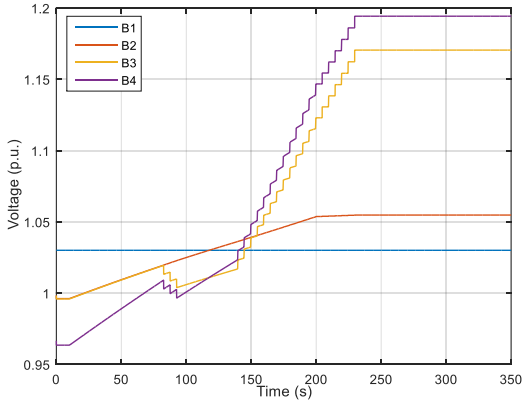
Fig. 8–(a). Feeder voltage profile: dynamic analysis.

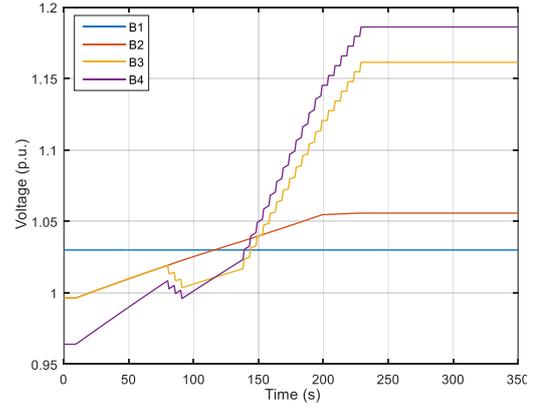
Fig. 8–(b). Feeder voltage profile: QSTS analysis.

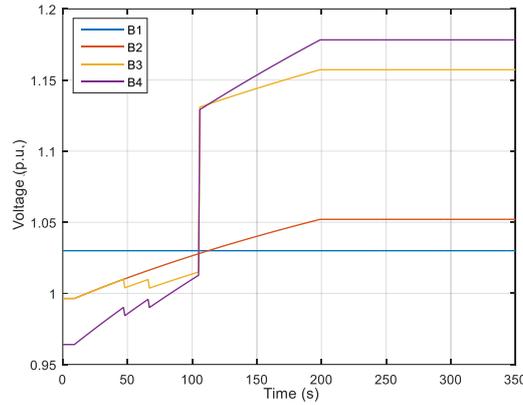
Fig. 8–(c). Feeder voltage profile: CLF analysis.

### 4.2 95-Bus UKGDS Test-Feeder

Fig. 9 shows the second test-feeder: the 95-bus United Kingdom generic distribution system (UKGDS) with 11 kV base voltage. This realistic radial network supplies four different kinds of consumer units – industrial, commercial, domestic unrestricted and domestic economy. The PS bus (bus 1) is the system's angular reference with 1 p.u. voltage setpoint. Runaway condition is observed in the SVR located between buses 23 and 24, with 0.98 p.u. voltage setpoint ($V_{ref}$ = 0.98 p.u.). The DG unit is set to operate in unity power factor control mode and static ZIP parameters are assigned to the distinct consumer load models accordingly.

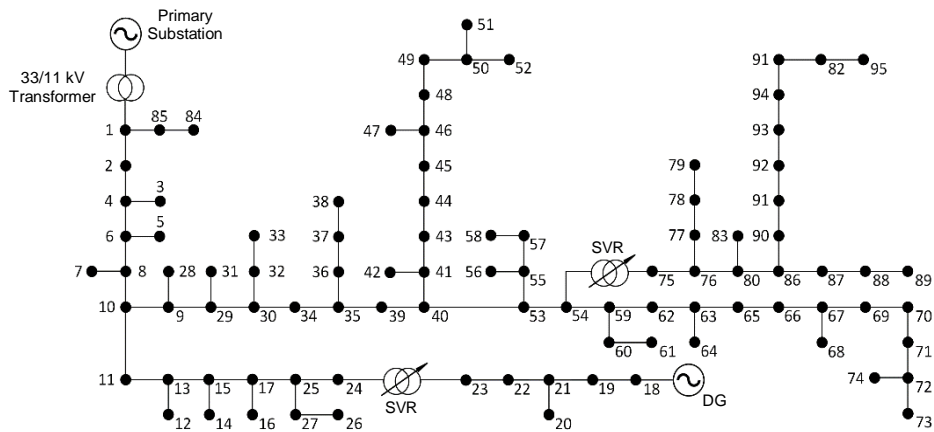
Fig. 9. Single-line diagram of the 95-bus UKGDS test-feeder. Source: adapted from [12].

Fig. 10 shows the active power flow variation through the SVR as the DG real power injection increases in linear fashion. Its ramp-up behavior starts at 10 s of simulation and stabilizes after reaching 2 MW at 200 s. Once DG penetration exceeds feeder loading, reverse active power flow through the SVR occurs (at around 95 s). This changes henceforth the regulation point of the bidirectional SVR from bus 23 to bus 24.

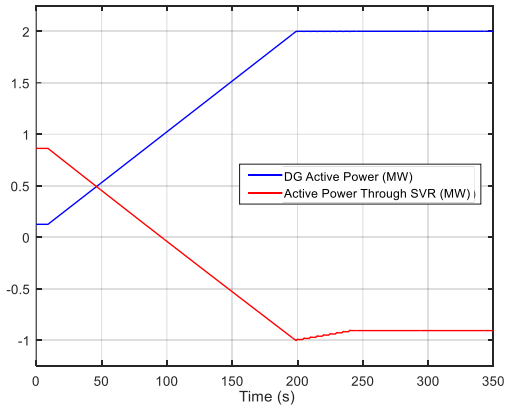

Fig. 10. DG active power ramp and active power flow through the SVR.

Fig. 11 compares the approaches in terms of SVR tap evolution and voltage profiles of SVR-regulated buses. During direct active power flow, the only disparity lies in the tap operation that happens exclusively with the CLF technique. During reverse active power flow, differences between simulations are accentuated as the upper tap limit is quickly reached in the CLF approach, leading to overvoltage at bus 23 which is further aggravated by the still ongoing DG ramp. However, the other simulations show that the upper tap limit is not reached until later, after DG injection stabilizes. This is another chronological discrepancy arising from time-dependent tap change delays. QSTS performance compared to the dynamic reference is once again adequate, with a small underestimation of bus 23 steady-state voltage (1.0974 p.u. against 1.1048 p.u., respectively).

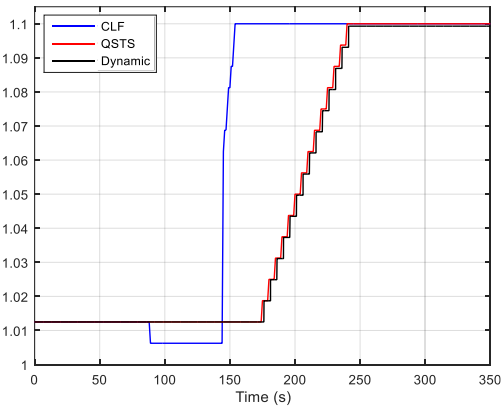
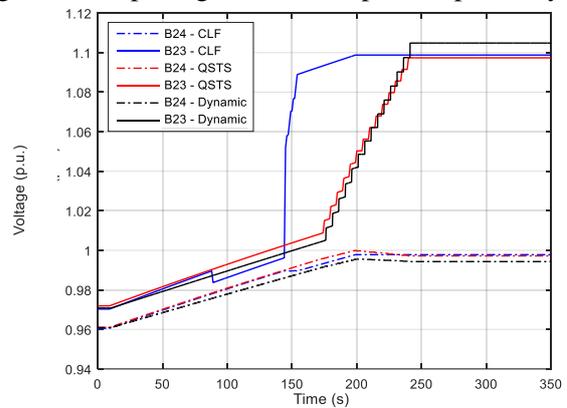

Fig. 11–(a). SVR tap evolution.  Fig. 11–(b). Voltage profiles of SVR-regulated buses.

Fig. 12 shows voltage profiles of buses located upstream and downstream of the SVR for each technique. All approaches are consistent in showing that the SVR is incapable of controlling the voltage of an upstream regulation point during reverse power flow, and that its failed attempts result in downstream overvoltage due to runaway condition. Nevertheless, only QSTS and dynamic simulations are able to provide an adequate reproduction of the sequence of events for the correct assessment of the phenomenon and its consequences.

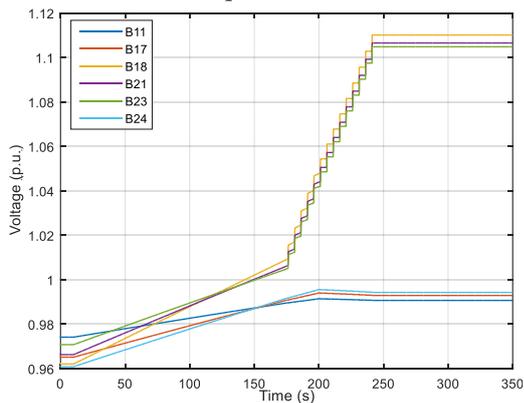
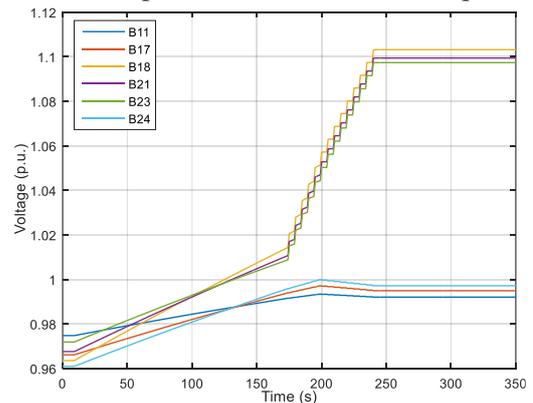

Fig. 12–(a). Feeder voltage profile: dynamic analysis.  Fig. 12–(b). Feeder voltage profile: QSTS analysis.

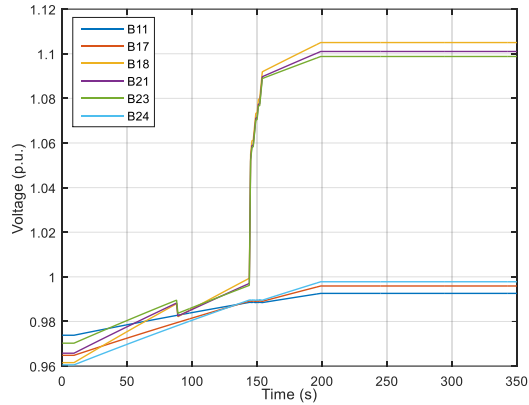

Fig. 12–(c). Feeder voltage profile: CLF analysis.

## 5  CONCLUSIONS

Results shown in this work highlight the need for chronological considerations when choosing the best simulation approach to assess SVR performance in a time window of minutes or seconds. The CLF method requires the least amount of computational effort, but is prone to temporal inconsistencies which might impair understanding of SVR runaway condition progression and its adverse effects. On the other hand, the dynamic method provides the closest approximation to a real scenario, but elevates computational effort as it employs overly detailed models too complex for the low-frequency dynamics inherent to SVR tap operations.

In this context, the QSTS method stands out as a simulation tool with low computational effort (6.9 ms of processing time against 6.6 ms with CLF approach and 1.83 s with dynamic approach for the UKGDS test-feeder) as well as high accuracy (0.77 % maximum underestimation error) for distribution network studies. It is, thus, an asset for DG interconnection impact assessments concerning SVR runaway condition chronology.

## 6  REFERENCES

[1] C.L. Masters, "Voltage rise: the big issue when connecting embedded generation to long 11 kV overhead lines", *Power Engineering Journal*, pp. 5-12, Aug. 2002.

[2] D. Ranamuka, et al., "Investigating the operation of multiple voltage regulators and DG in a distribution feeder", in *International Conference on Advances in Energy Engineering*, Bangkok, Dec. 2011.

[3] H.R. Brito, V.M. Souza, V.C. Souza, J.P.A. Vieira, et al., "Impact of distributed generation on distribution systems with cascaded bidirectional step voltage regulators", in *International Conference on Industry Applications*, Sao Paulo, Nov. 2018.

[4] I. Dzafic, R.A. Jabr, E. Halilovic, and B.C. Pal, "A sensitivity approach to model local voltage controllers in distribution networks", *IEEE Trans. on Power Systems*, pp. 1419-1428, May 2014.

[5] J.A.D. Massignan, B.R. Pereira Jr, et al., "Load flow calculation with voltage regulators bidirectional mode and distributed generation", *IEEE Trans. on Power Systems*, pp. 1576-1577, Jun. 2016.

[6] D. Ranamuka, A.P. Agalgaonkar, and K.M. Muttaqi, "Online voltage control in distribution systems with multiple voltage regulating devices", *IEEE Trans. on Sustainable Energy*, pp. 617-628, Sept. 2013.

[7] C.A. Colopy, S. Grimes, and J.D. Foster, "Proper operation of step voltage regulators in the presence of embedded generation", in *CIRED Conference*, Nice, Jun. 1999.

[8] J.E. Quiroz, M.J. Reno, and R.J. Broderick, "Time series simulation of voltage regulation device control modes", in *Photovoltaic Specialists Conference*, Tampa, Jun. 2013.

[9] A. Khoshkbar-Sadigh, et al., "The necessity of time-series simulation for investigation of large-scale solar energy penetration", in *Innovative Smart Grid Technologies Conference*, Washington, Feb. 2015.

[10] D. Paradis, F. Katiraei, and B. Mather, "Comparative analysis of time-series studies and transient simulations for impact assessment of PV integration on reduced IEEE 8500 node feeder", in *IEEE PES General Meeting*, Vancouver, Jul. 2013.

[11] P. Bagheri, Y. Liu, W. Xu, et al., "Mitigation of DER-caused over-voltage in MV distribution systems using voltage regulators", *IEEE Power and Energy Technology Systems Journal*, in press, Nov. 2018.

[12] V.M. Souza, H.R. Brito, et al., "QSTS simulation of reverse power tap changer runaway condition in active distribution systems", in *Brazilian Symposium on Power Systems*, Niteroi, May 2018.